\documentclass[12pt,article]{article}
\usepackage{fancyhdr}

\usepackage{amsmath,amssymb}
\usepackage{physics}

\usepackage{mathtools}
\usepackage{graphicx}
\usepackage{color}

\begin{document}

	\date{}
	
	\title{\Large\bf Air Flow Analysis of a Rotating Cylinder through Numerical Simulation }
	
	\author{ Alan Hsu\textsuperscript{1} and Fei Liu\textsuperscript{2} \\
		\textsuperscript{1} \textit{John P. Stevens High School, Edison, NJ, 08820,  email: alanhsu.1@gmail.com} \\
		\textsuperscript{2} \textit{New Jersey Science Academy, Fei.Liu@njsci.org}
		}

	\maketitle

	\section*{\centering Abstract}

	\textit{The complete flow field surrounding a rotating cylinder is calculated by solving the Navier-Stokes equations using the finite difference method. The numerical simulation is performed on a transformed rectilinear grid, with axes representing the radial and angular dimensions. Different boundary conditions of the simulations are tested by changing the tangential speed of the rotating cylinder, ranging from subsonic to supersonic speed. The following variables in the modeled flow fields are analyzed: temperature, velocity, pressure, density, and stress. The large gradient of velocity near the cylinder is found to create high stress, which leads to rise in temperature. Pressure and density near the cylinder decreases as the speed of the cylinder increases. The proposed simulation method and analysis of air flow can be extended to the modeling of air flow around a Diabolo.}

	\section{Introduction}
	Computational Fluid Dynamics (CFD) is developed to solve diverse fluid mechanics problems~\cite{Anderson} with various numerical techniques such as finite difference ~\cite{Richardson}~\cite{Ghadimi}, finite volume ~\cite{Hirsch}~\cite{Moukalled}~\cite{Versteeg}, and finite element~\cite{Zienkiewicz}~\cite{Hosain} methods. These techniques have been used to find solutions to the driven-lid cavity problem~\cite{Matyka}, low Reynolds number flow (Re = 40)
	about a circular cylinder on an unstructured adaptive grid made of triangles and rectangles~\cite{Holmes}, and the downstream flow field surrounding a 2D rectangular prism~\cite{Yu}.
	
	In this paper, the Navier-stokes equations are solved to investigate the air flow properties around a cylinder rotating about its central axis. Past research on air flow around a cylinder includes the affect of surface roughness on the cylinder~\cite{Merrick}, the affect of a surrounding porous medium on the air~\cite{Sobera}, the difference in flow fields between a circular and square cylinder~\cite{Gera}, and the turbulent air flow of a high Reynolds number around a cylindrical body~\cite{Pang}. Most of the discussions in the past have been focusing on downstream air flow around a stationary cylinder. Another interesting research topic is the flow field generated from a cylinder spinning on its central axis, which can be extended to real-world spinning objects such as a Diabolo or wheels of a car.

	\section{Problem and Proposed Airflow model}
	
\begin{figure}[!htb]
		\begin{minipage}[b]{1.0\linewidth}
			\includegraphics[width=18cm]{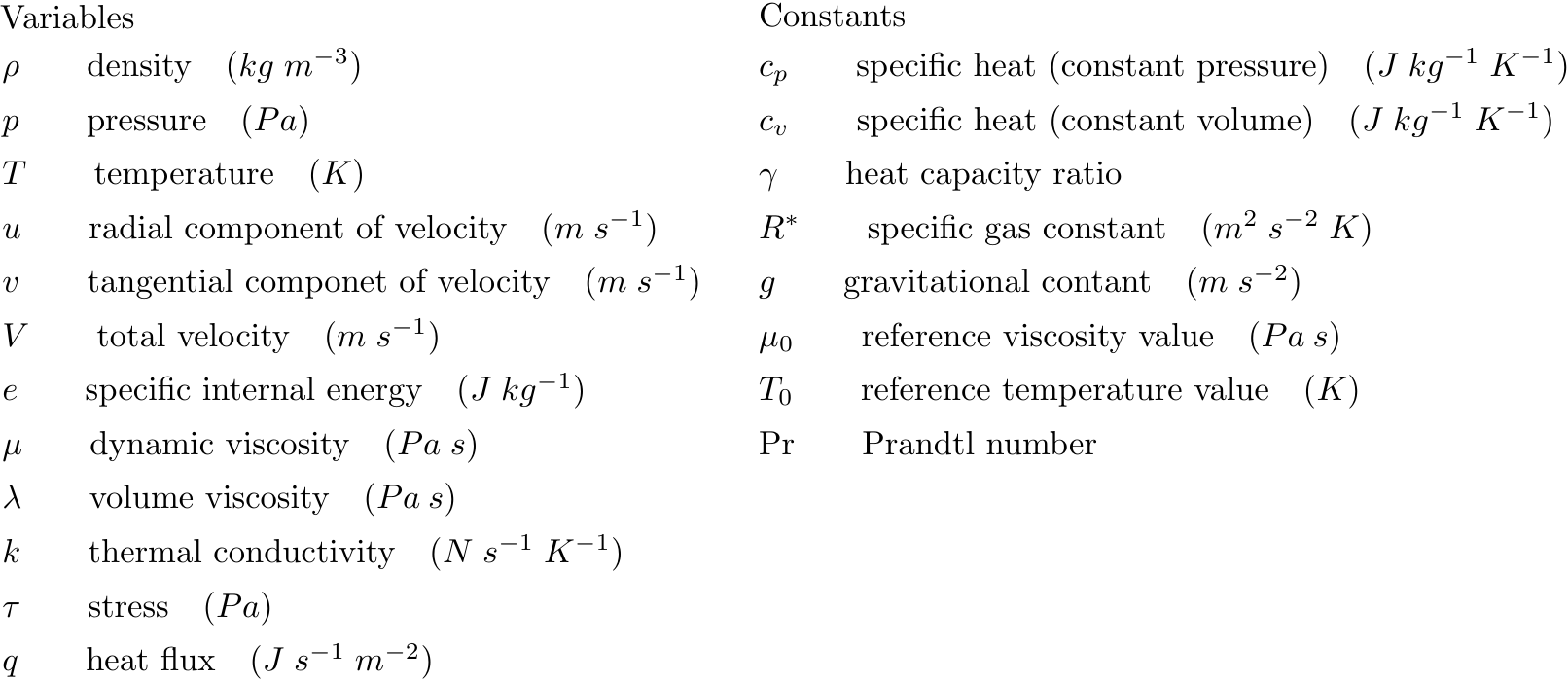}
		\end{minipage}
		\caption{Variables Used in the Simulation.}
		\label{fig:VariablesConstants}
	\end{figure}

	The given problem is to find the solution to the air flow surrounding a rotating cylinder under various boundary conditions and parameters. The Navier-Stokes equations are used to solve the air flow which is 2-D, compressible, viscous, and laminar. The behavior of the air flow using different boundary conditions such as speed is examined. In the model, we define the cylinder of radius $r_0$ to rotate along its central axis with a tangential velocity $v_{tan}$, and a solution to the complete flow field for a certain $r_1$ distance from the surface of the cylinder is produced. In the equations, the variables $u$, $v$, $\rho$, and $e$ denote radial velocity, tangential velocity, density, and specific internal energy, respectively, as shown below with all the other variables and constants in Figure~\ref{fig:VariablesConstants}. In the model, the physical configuration on a polar grid is transformed to Cartesian parametric grid.

	\subsection{Governing Equations}
	To solve for the flow field, we are going to solve the 2-D Navier Stokes equations. Before grid transformation, the Navier-Stokes equations is given by: 
	
	\begin{equation} \label{eq:1}
	\pdv{U}{t}+\pdv{F}{x}+\pdv{G}{y}=0,
	\end{equation} 
	
	Where $U$, $F$, and $G$ are following solution vectors: 
	\begin{equation} \label{eq:2}
	U =\left( \begin{array}{clcr}
	\rho \\
	\rho \; u \\
	\rho \; v \\
	\rho \; (e + \frac{V^2}{2}) 
	\end{array} \right),
	\end{equation}
	
	\begin{equation} \label{eq:3}
	F =\left( \begin{array}{clcr}
	\rho \; u\\
	\rho \; u^2 + p - \tau_{xx} \\
	\rho \; u \; v - \tau_{xy} \\
	(\rho \; (e+\frac{V^2}{2}) + p) \; u - u \; \tau_{xx} - v \; \tau_{xy} + q_x \\
	\end{array} \right),
	\end{equation}
	
	\begin{equation} \label{eq:4}
	G =\left( \begin{array}{clcr}
	\rho \; v\\
	\rho \; u \; v - \tau_{xy} \\
	\rho \; v^2 + p - \tau_{yy} \\
	(\rho \; (e+\frac{V^2}{2}) + p)  \; v - u \; \tau_{xy} - v \; \tau_{yy} + q_y \\
	\end{array} \right).
	\end{equation}
	
	Equations~\ref{eq:1} -~\ref{eq:4} represent the two dimensional conservative form of the Navier-Stokes equations. In order to completely solve the fluid equations above, a few other conditions need to be included:
	
	\begin{equation*} \label{eq:5}
	p = \rho \; R^* \;  T , \text{where } R^* \text{ is the gas constant per molar mass}. 
	\end{equation*}
	\begin{equation*} \label{eq:6}
	e = \frac{1}{\gamma - 1} \; \frac{p}{\rho}, \text{where } \gamma = \frac{c_p}{c_v} \text{. For air, the value of } \gamma \text{ is around 1.4.} 
	\end{equation*}
	\begin{equation*} \label{eq:7}
	V = \sqrt{u^2+v^2},
	\end{equation*}
	\begin{equation*} \label{eq:8}
	\mu=\mu_0 \; (\frac{T}{T_0})^{\frac{3}{2}} \; (\frac{T_0+110}{T+110}), \text{ which is also known as Sutherland's law.} 
	\end{equation*}
	\begin{equation*} \label{eq:9}
	\lambda=-\frac{2}{3}\; \mu,
	\end{equation*}
	\begin{equation*} \label{eq:10}
	\frac{\mu \; c_p}{k}=0.71, \text{ which is also known as the Prandtl number.} \\
	\end{equation*}
	
	We will also need the definitions of the stress ($\tau_{xy}, \tau_{xx}, \tau_{yy}$) and heat flux ($q_x, q_y$): 
	\begin{equation} \label{eq:11}
	\tau_{xy}=\tau_{yx}=\mu \;  (\pdv{u}{y}+\pdv{v}{x}),
	\end{equation}
	\begin{equation} \label{eq:12}
	\tau_{xx}=\lambda \; (\grad{\cdot \vec{V}}) + 2 \mu \; \pdv{u}{x},
	\end{equation}
	\begin{equation} \label{eq:13}
	\tau_{yy}=\lambda \; (\grad{\cdot \vec{V}}) + 2 \mu \; \pdv{v}{y}, 
	\end{equation}
	\begin{equation} \label{eq:14}
	q_x=-k\; \pdv{T}{x}, 
	\end{equation}
	\begin{equation} \label{eq:15}
	q_y=-k\; \pdv{T}{y}. 
	\end{equation}
	
	\subsection{Grid Transformation Equations}
	Since the cross-section of the cylinder is circular, it is natural to use a polar coordinate system. To facilitate finite difference calculation, the set of differential Eqs. (\ref{eq:1}-\ref{eq:4}) is transformed from the polar coordinate system to the Cartesian coordinate system. Let the radius $\xi$ be $r_0 +r$ and the polar angle $\eta$ be $\theta$. Then, the Cartesian position coordinates are $x=\xi \: cos(\eta)$ and $y=\xi \:  sin(\eta)$. Based on the coordinate transformation, elements of the Jacobian are: 
	\begin{equation*} \label{eq:16}
	\pdv{y}{\eta}=\xi \; cos(\eta),
	\end{equation*}
	\begin{equation*} \label{eq:17}
	\pdv{y}{\xi}=sin(\eta),
	\end{equation*}
	\begin{equation*} \label{eq:18}
	\pdv{x}{\eta}=-\xi \; sin(\eta),
	\end{equation*}
	\begin{equation*} \label{eq:19}
	\pdv{x}{\xi}=cos(\eta).
	\end{equation*}

	The Jacobian of the $x,y,\xi,\eta$ vector transformation is defined as:
	\begin{equation*} \label{eq:20}
	\begin{bmatrix}
	\pdv{x}{\xi} & \pdv{y}{\xi} \\
	\pdv{x}{\eta} & \pdv{y}{\eta} \\
	\end{bmatrix}.
	\end{equation*}
	
	The matrices in Equation~\ref{eq:1} are transformed in Eqs. (\ref{eq:21}-\ref{eq:23}) to corresponding matrices in the computational plane:
	
	\begin{equation} \label{eq:21}
	U_1=J\; U,
	\end{equation}
	\begin{equation} \label{eq:22}
	F_1=F \; \pdv{y}{\eta} - G \; \pdv{x}{\eta}, 
	\end{equation}
	\begin{equation} \label{eq:23}
	G_1=-F \; \pdv{y}{\xi}+G \; \pdv{x}{\xi}. 
	\end{equation}

	The following two differential operators are used to transform any partial differentials of variables mentioned in Eqs. (\ref{eq:11}-\ref{eq:15}) from the physical domain to the computational domain:
	\begin{equation} \label{eq:24}
	\pdv{}{x}=\frac{1}{J} \left( \pdv{}{\xi} \pdv{y}{\eta} - \pdv{}{\eta} \pdv{y}{\xi} \right),
	\end{equation}
	\begin{equation} \label{eq:25}
	\pdv{}{y}=\frac{1}{J} \left( \pdv{}{\eta} \pdv{x}{\xi} - \pdv{}{\xi} \pdv{x}{\eta} \right).
	\end{equation}
	
	The Navier-Stokes equations solved over the computational domain is 
	\begin{equation} \label{eq:26}
	\pdv{U_1}{t}+\pdv{F_1}{x}+\pdv{G_1}{y}=0.
	\end{equation}
	
	\subsection{Initial and Boundary Conditions}
	
	\begin{figure}[!htb]
	\begin{minipage}[b]{1.0\linewidth}
		\centering
		
		\includegraphics[width=16cm]{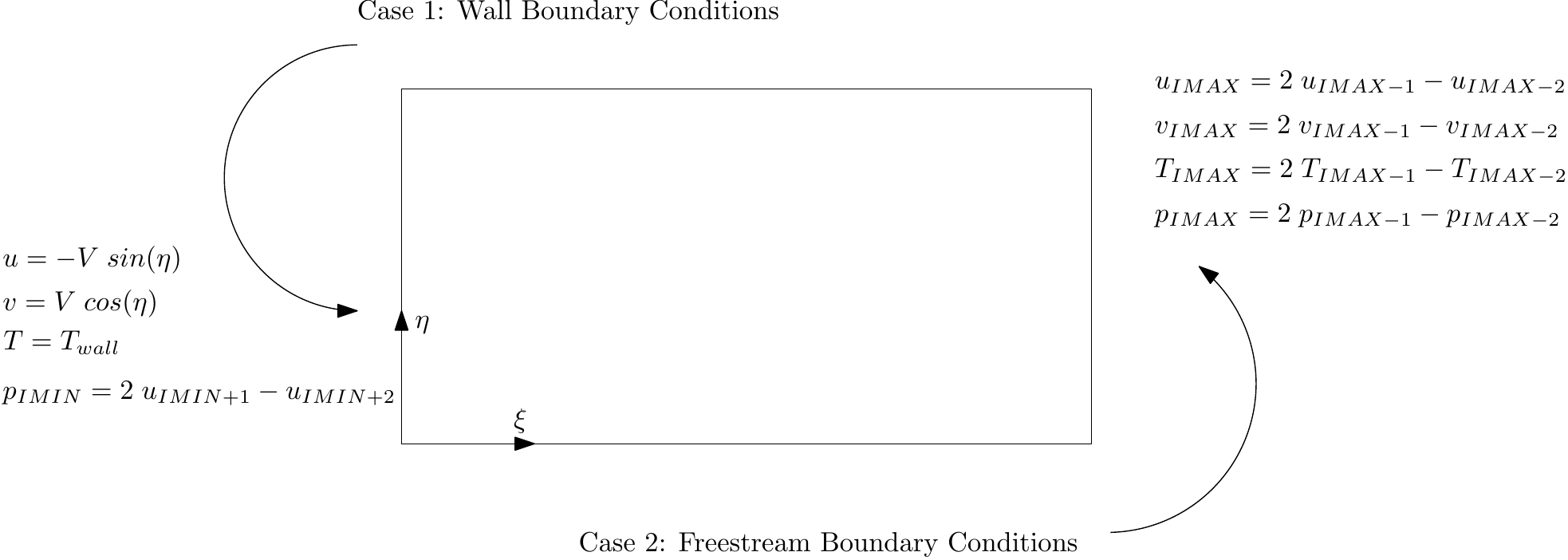} 
		
	\end{minipage}
	\caption{Boundary Conditions.}
	\label{fig:BoundaryConditions}
\end{figure}
	
	Second order Maccormack~\cite{Maccormack} algorithm in both space and time domain with explicit finite difference is used to solve the Navier-Stokes Equations. The computational plane has a radial axis $\xi$ and an angular axis $\eta$. In the $\eta$ direction, there is no outer or inner boundary because of its periodic nature; however, in the $\xi$ (radial) direction, the surface of the cylinder is at one side and the free-stream boundary is at the other. In other words, the wall is at $\xi=r_0$ and the free-stream boundary is at $\xi=r_0+r_{max}$. 
	
	For the initial conditions, the temperature is 300 Kelvin, the $u$ and $v$ components of velocity are 0 m/s except at the wall boundary, the pressure is 101.325 kPa, and the stress and heat flux terms are all 0. For the boundary conditions shown in Figure~\ref{fig:BoundaryConditions}, the temperature at the wall is always 300 Kelvin, and is extrapolated at the free-stream boundary. The velocity at the wall, $v_{tan}$, is set to various speeds, and the free-stream values of $u$ and $v$ components are extrapolated. Pressure is extrapolated both at the wall and the free-stream boundary. 
	
	The initial and boundary conditions for density, internal energy, dynamic viscosity, volume viscosity, and thermal conductivity are then calculated from the existing variables.
	
	\section{Numerical Method}

	The rectilinear computational $\xi,\eta$ grid is constructed with appropriate step sizes, $2 \times 10^{-7} \; m$ in the $\xi$ direction and $0.009 \; m$ in the $\eta$ direction. The Reynolds number is set to various values for the simulation. To satisfy the CFL condition, subsonic flows use a time-step around $10^{-11} \; s$ to prevent instability. For supersonic flows, a smaller time step $10^{-12} \; s$ is used.
	
	The transformation from $x,y$ to $\xi,\eta$ is performed to generate the necessary partial derivatives to calculate the Jacobian for later use. In order to maintain second order accuracy in the spatial domain, the difference leading to $\pdv{U}{t}$ is calculated in opposite direction during corrector and predictor steps~\cite{Maccormack}. 
	
	During the predictor step, a forward difference is used for the $F$ and $G$ calculation, as shown in Eq. (\ref{eq:U1pred}):
	\begin{equation}\label{eq:U1pred}
	\pdv{U_1}{t}_{Pred}=-(\frac{F_{1(i+1,j)}-F_{1(i,j)}}{d \xi}+\frac{G_{1(i,j+1)}-G_{1(i,j)}}{d \eta}).
	\end{equation}

	A partial derivative in Eq. (\ref{eq:U1pred}) used in the matrix of $F$ with respect to $\xi$ will be differenced backward, while one with respect to $\eta$ will be differenced centrally. Similarly, a partial derivative in Eq. (\ref{eq:U1pred}) in the matrix $G$ with respect to $\eta$ will be differenced backward, while one with respect to $\xi$ will be differenced centrally.
	
	$F$ and $G$ matrices are then constructed with corresponding vector-transformations to get the $F_1$ and $G_1$ necessary for the predictor step, shown in Eqs. (\ref{eq:21}-\ref{eq:23}). Then, $\pdv{U}{t}_{predictor}$ is calculated by forward differencing $F_1$ and $G_1$, where $F$ is differenced with respect to $\xi$ and $G$ is differenced with respect to $\eta$. Using $F_1$ and $G_1$, the intermediate value of $U_1$ is generated.
	
	Before the corrector step, $U_1$ must be decoded to obtain the data to get the primitive variables in order to generate new functions for $\bar{F}$ and $\bar{G}$. After updating the primitive variables, the variables are extrapolated at the free-stream boundary, shown in Figure~\ref{fig:BoundaryConditions}. Finally, the heat and friction partial derivative terms are also updated for the corrector step. 
	
	During the corrector step, a backward difference is used for the F and G calculation. A partial derivative in Eq. (\ref{eq:U1Corr}) used in the matrix of $F$ with respect to $\xi$ will be differenced forward, while one with respect to $\eta$ will be differenced centrally. Similarly, a partial derivative in Eq. (\ref{eq:U1Corr}) in the matrix $G$ with respect to $\eta$ will be differenced forward, while one with respect to $\xi$ will be differenced centrally. Now that all of the data is updated, $\bar{F}$ and $\bar{G}$ are constructed for the corrector step, shown in Eq. (\ref{eq:U1Corr}):

	\begin{equation} \label{eq:U1Corr}
	\pdv{U_1}{t}_{Corr} = - (\frac{\bar{F}_{1(i,j)}-\bar{F}_{1(i-1,j)}}{d \xi} +\frac{\bar{G}_{1(i,j)}-\bar{G}_{1(i,j-1)}}{d \eta}),
	\end{equation}
	
	$\pdv{U}{t}_{corrector}$ is then calculated by backward differencing $\bar{F}$ and $\bar{G}$. 
	
	\begin{equation} \label{eq:U1}
	\pdv{U_1}{t}^n = \frac{1}{2}( \pdv{U_1}{t}_{pred} + \pdv{U_1}{t}_{corr}),
	\end{equation}
	
	\begin{equation} \label{eq:updateU}
	U_1^{n+1} = U_1^n + \pdv{U_1}{t}^n dt.
	\end{equation}
	
	Finally, the value of $U_1$ at time step $n+1$ is updated by taking the $U_1$ at time step $n$ and adding the average of the predictor and corrector values as described by Eqs. (\ref{eq:U1}-\ref{eq:updateU}). A similar process of decoding is done to obtain the new values of the variables such as $\rho$, $p$, $T$,  ready for the next iteration.

	\section{Simulation Results and Discussion}
	The simulation is run until the disturbance in the flow field reaches the outer boundary. The distribution of 5 variables from subsonic and supersonic simulations are examined: temperature, density, pressure, velocity, and stress. 
    
\begin{figure}[!h]
		\begin{minipage}[b]{0.45\linewidth}
			\centering
			
			\includegraphics[width=9cm]{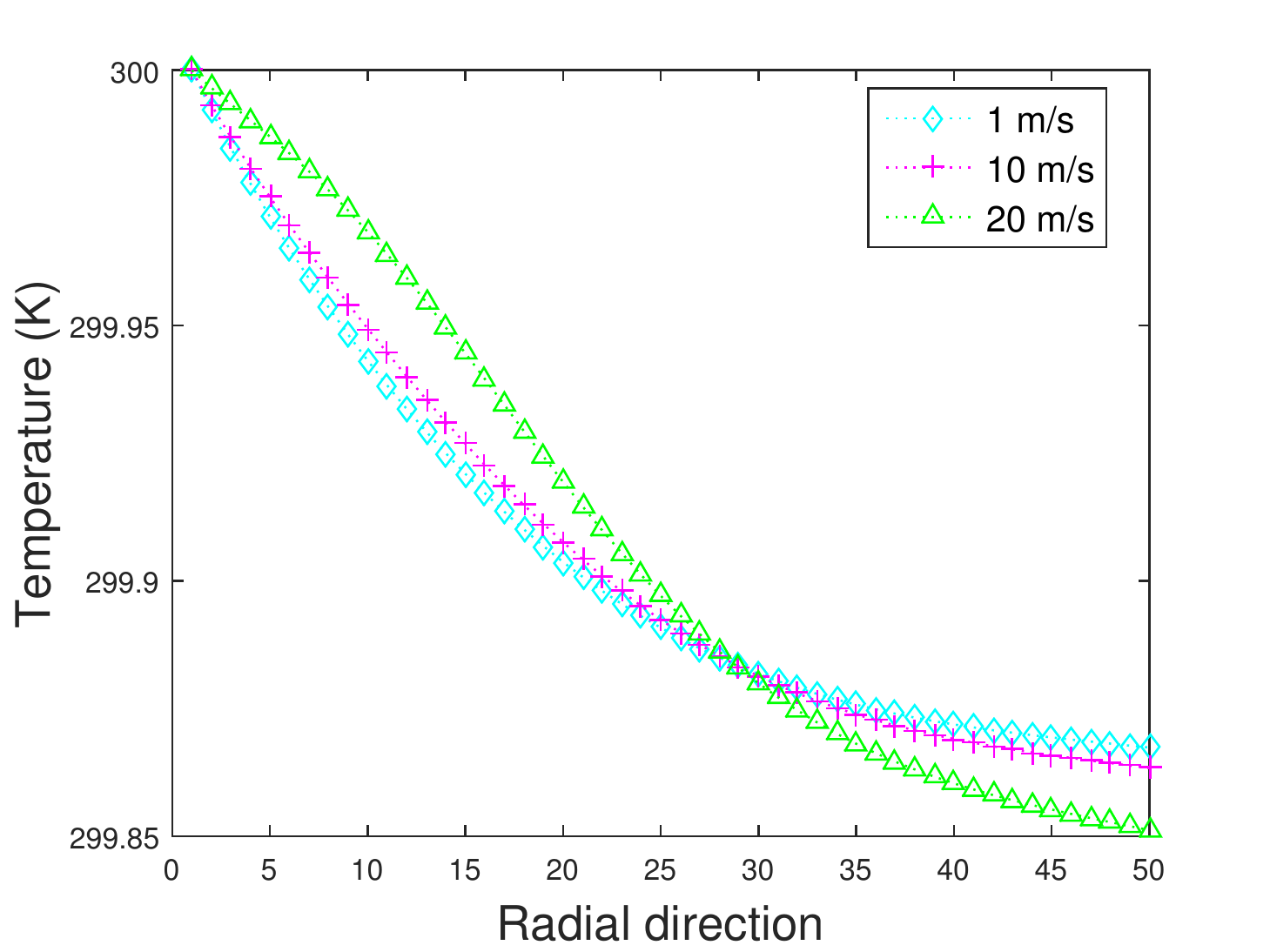}
			
			\caption{Temperature Distribution at Subsonic Speeds.}
			\label{fig:TempSub}
		\end{minipage}
		\hspace{1cm}
		\begin{minipage}[b]{0.4\linewidth}
			\centering
			
			\includegraphics[width=9cm]{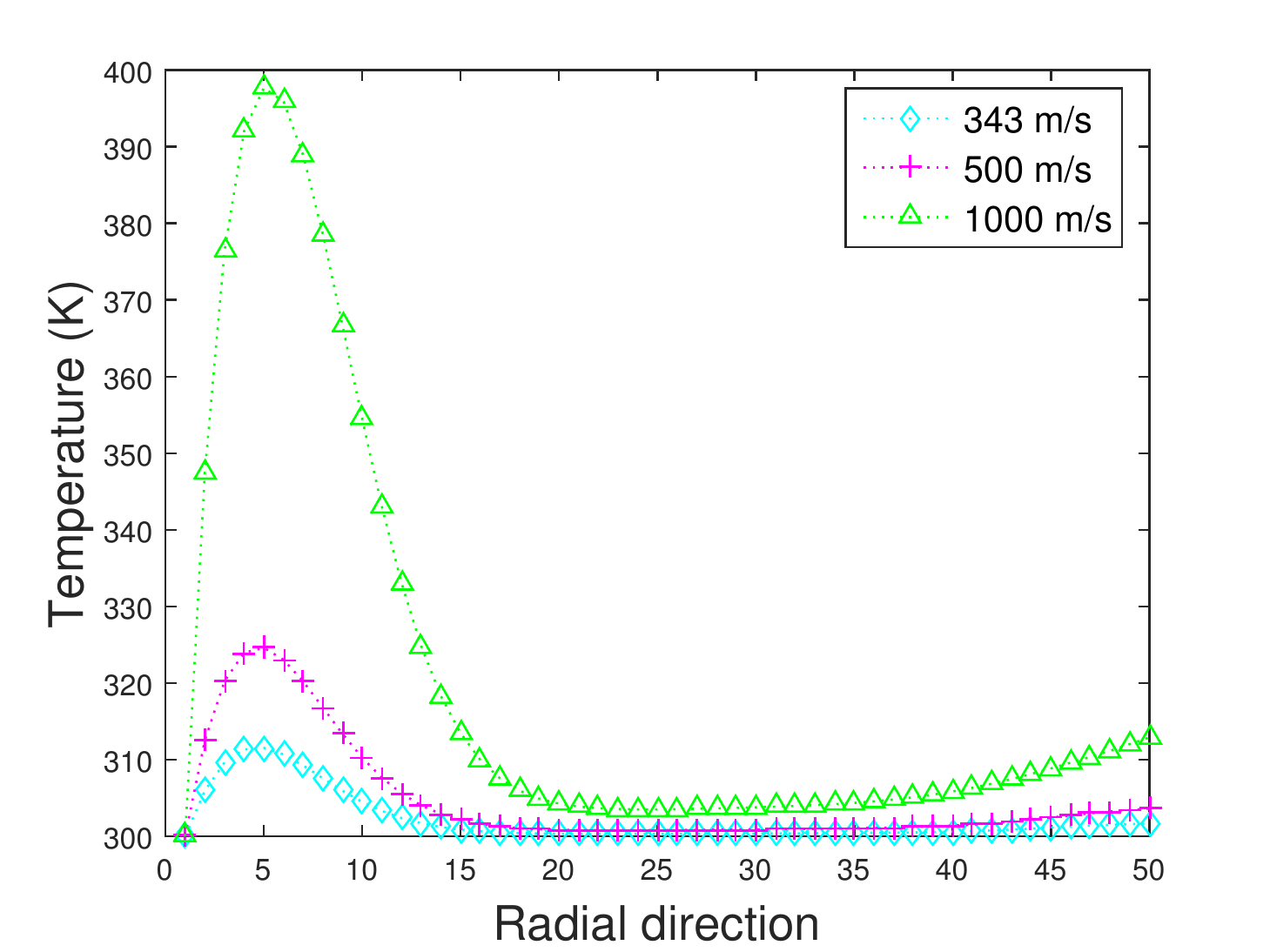}
			
			\caption{Temperature Distribution at Supersonic Speeds.}
			\label{fig:TempSuper}
		\end{minipage}
	\end{figure}

\begin{figure}[!h]
		\begin{minipage}[b]{0.45\linewidth}
			\centering
			
			\includegraphics[width=9cm]{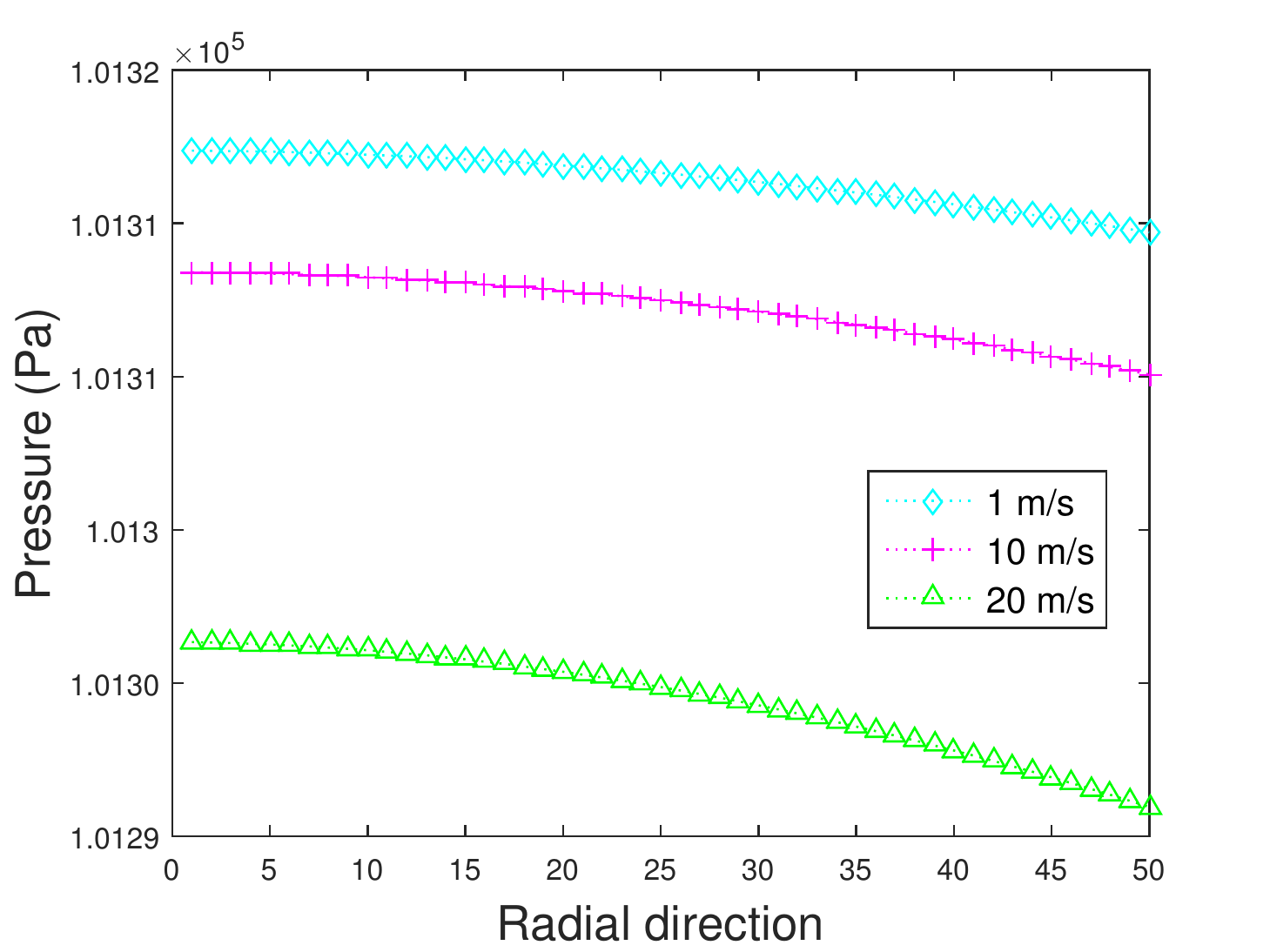}
			
			\caption{Pressure Distribution at Subsonic Speeds.}
			\label{fig:PressureSub}
		\end{minipage}
		\hspace{1cm}
		\begin{minipage}[b]{0.45\linewidth}
			\centering
			
			\includegraphics[width=9cm]{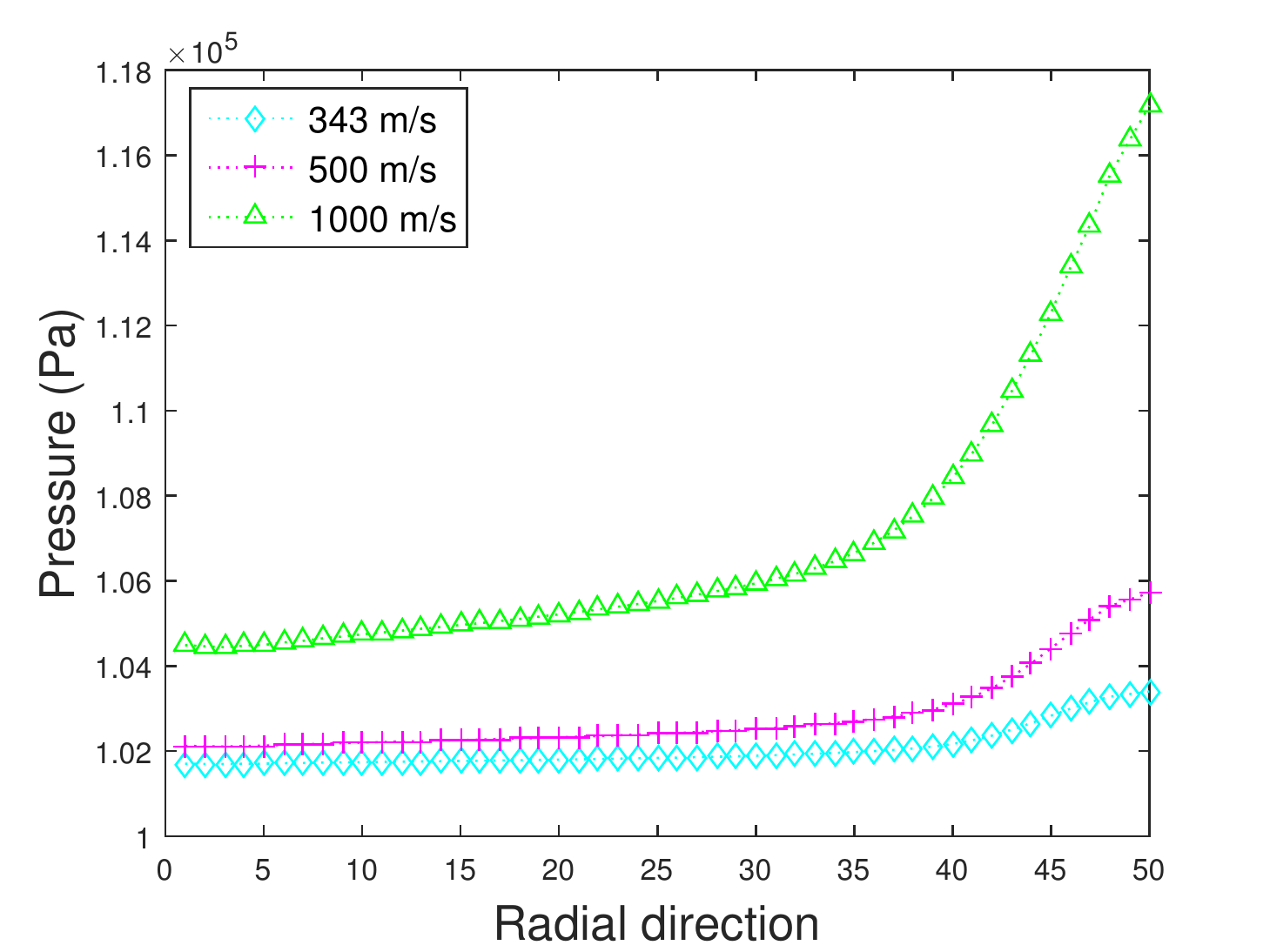}
			
			\caption{Pressure Distribution at Supersonic Speeds.}
			\label{fig:PressureSuper}
		\end{minipage}
	\end{figure}

\begin{figure}[!h]
		\begin{minipage}[b]{0.45\linewidth}
			\centering
			
			\includegraphics[width=9cm]{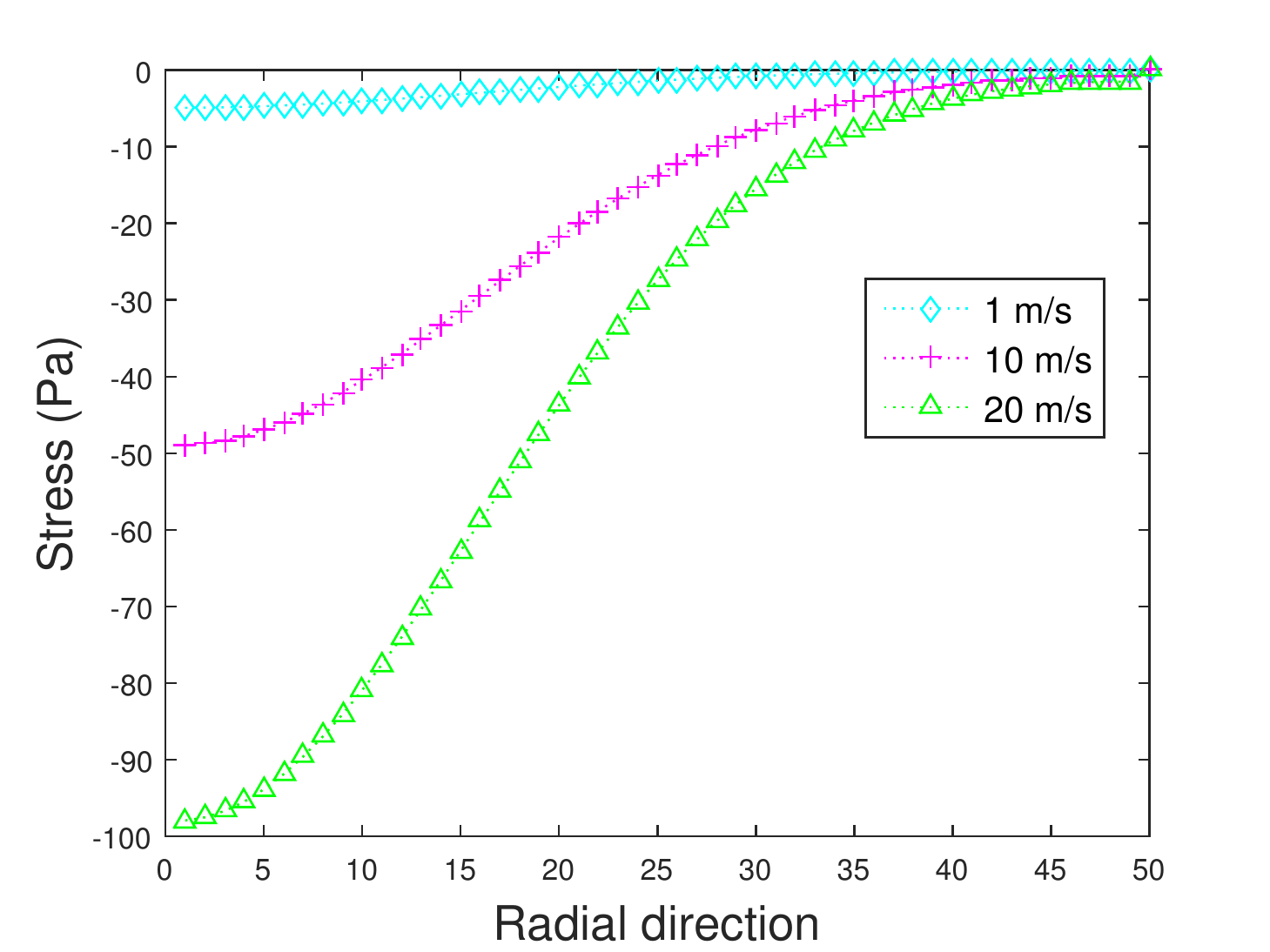}
			
			\caption{Stress Distribution at Subsonic Speeds.}
			\label{fig:StressSub}
		\end{minipage}
		\hspace{1cm}
		\begin{minipage}[b]{0.45\linewidth}
			\centering
			
			\includegraphics[width=9cm]{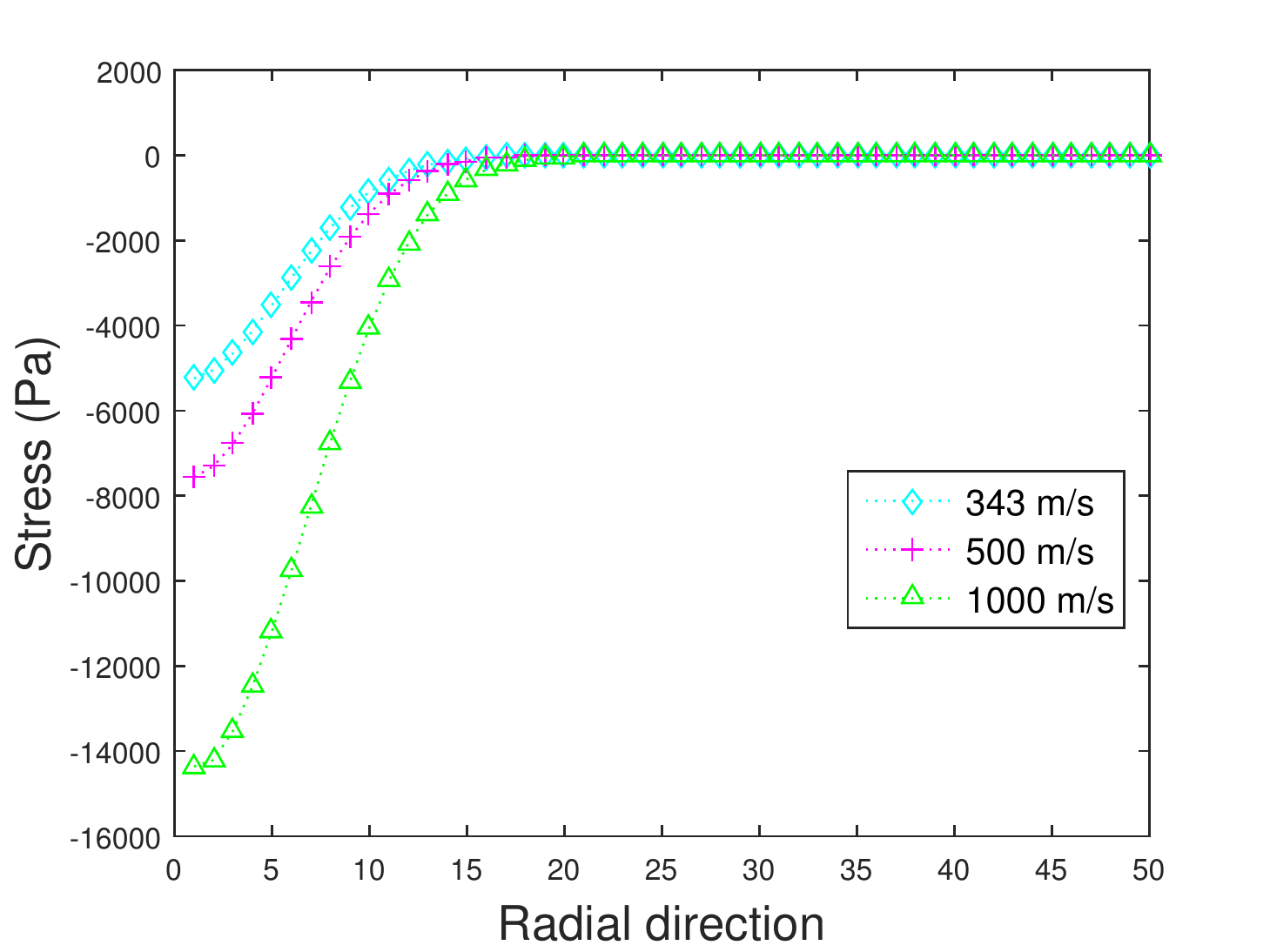}
			
			\caption{Stress Distribution at Supersonic Speeds.}
			\label{fig:StressSuper}
		\end{minipage}
	\end{figure}

    Near the cylinder wall, the behavior of temperature differs between low and high speeds. In Figure~\ref{fig:TempSub} and~\ref{fig:TempSuper}, temperature monotonically decreases radially outward. However, the distributions of 1 m/s and 10 m/s starts out concave, while the 20 m/s starts out convex. The convex graph is more representative in supersonic results, where the temperature rapidly increases and then decreases near the wall of the cylinder. For 1000 m/s, the maximum temperature approaches 400K, which is consistent with the large frictional stress near the cylinder as shown in Figure~\ref{fig:StressSub}. On the other hand, near the free-stream boundary where the friction is essentially zero shown in Figures~\ref{fig:StressSub} and~\ref{fig:StressSuper}, temperature is primarily determined by pressure for both subsonic and supersonic results. Temperature increases as pressure increases according to the ideal gas law, shown in Figures~\ref{fig:PressureSub} and~\ref{fig:PressureSuper}.
    
\begin{figure}[!h]
		\begin{minipage}[b]{0.45\linewidth}
			\centering
		
			\includegraphics[width=9cm]{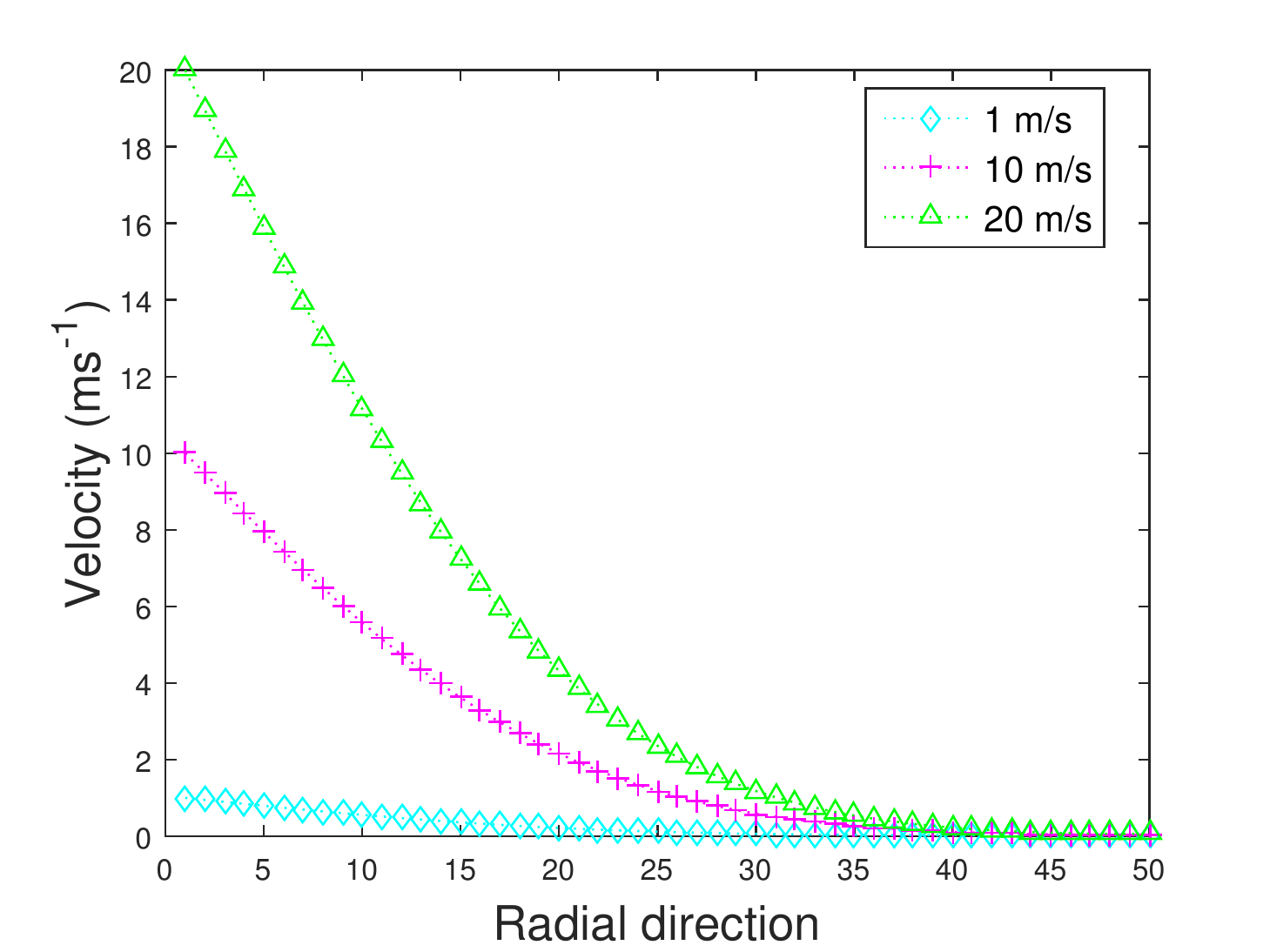}
			
			\caption{Velocity Distribution at Subsonic Speeds.}
			\label{fig:VSub}
		\end{minipage}
		\hspace{1cm}
		\begin{minipage}[b]{0.45\linewidth}
			\centering
			
			\includegraphics[width=9cm]{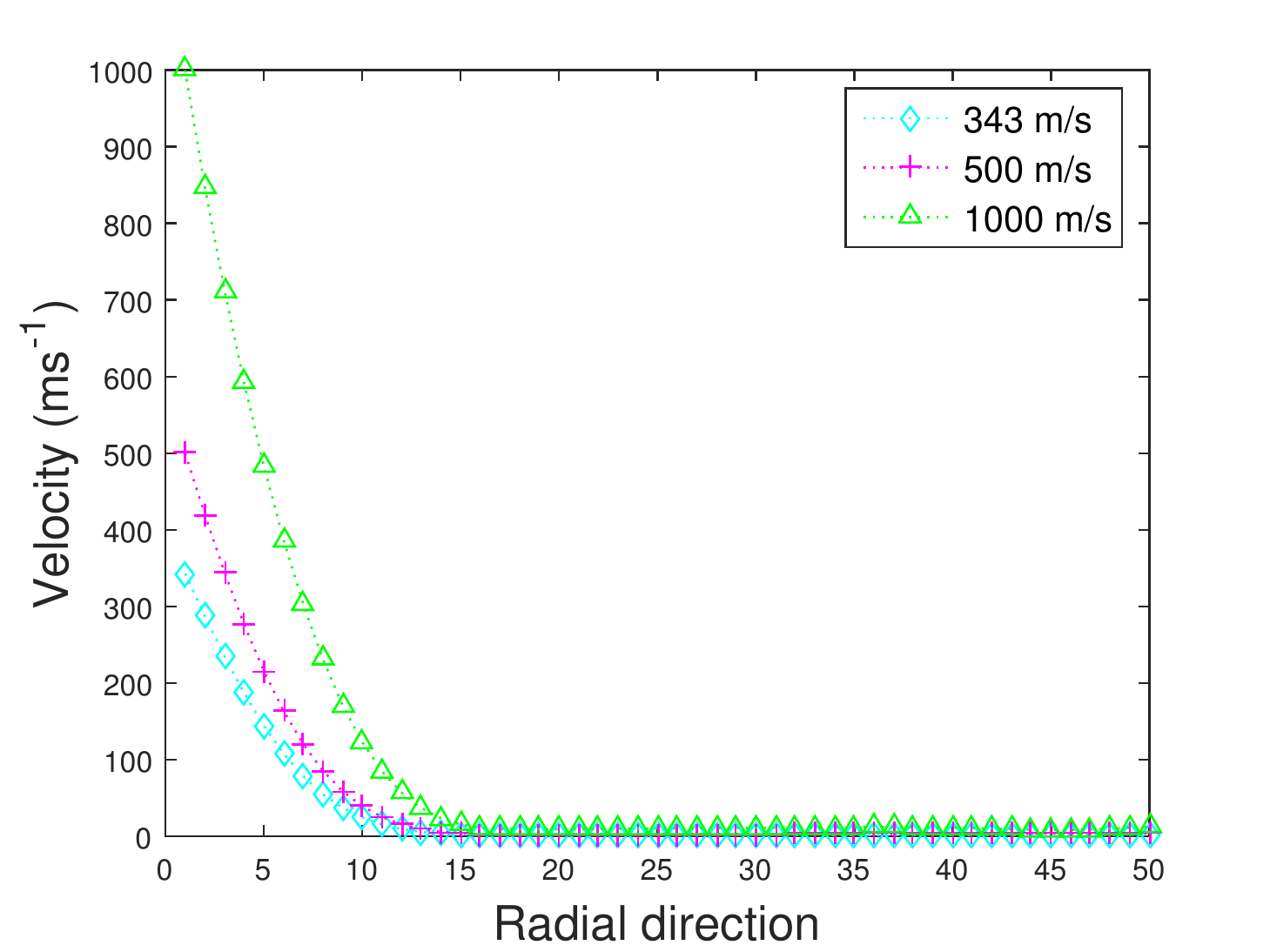}
			
			\caption{Velocity Distribution at Supersonic Speeds.}
			\label{fig:VSuper}
		\end{minipage}
	\end{figure}

    The velocity at low speeds converges to zero around 40 units radially away from the cylinder, as shown in Figure~\ref{fig:VSub}, while only at 15 units for supersonic speeds shown in Figure~\ref{fig:VSuper}. This is consistent with the stress ($\tau$ terms), shown in shown in Figures~\ref{fig:StressSub} and~\ref{fig:StressSuper}. The stresses graphed here are the mixed partial derivatives of the $u$ and $v$ velocity. Because of the high gradients of velocity near the wall of the cylinder, there is high stress near the wall as well. At 1000 m/s, the stress can reach over 14000 Pascals as shown in the green curve of Figure~\ref{fig:StressSuper}, which is consistent with the 400 K temperature at that point as well. Conversely, the velocity near the free-stream boundary tend to zero because of the no-slip boundary condition present in both subsonic and supersonic simulations. Likewise, the stress must also tend to zero near the free-stream since it is defined as the gradient of velocity.
    
    At subsonic speeds, the data for pressure is consistent with the velocity. By Bernoulli's Principle, high velocities will result in low pressures, and low velocities result in high pressure, as shown in Figure~\ref{fig:PressureSub}. With the largest velocity (20 m/s), the pressure is at the lowest $(1.0130 \times 10^5)$. In supersonic cases, the Bernoulli effect is more noticeable , shown in Figure~\ref{fig:PressureSuper}. Near the cylinder wall, high speeds faster than the speed of sound result in very low pressure. Near the free-stream boundary, as velocities tend to zero, pressure increases, relatively quickly in the 1000 m/s case. 
    
\begin{figure}[!h]
		\begin{minipage}[b]{0.45\linewidth}
			\centering
			
			\includegraphics[width=9cm]{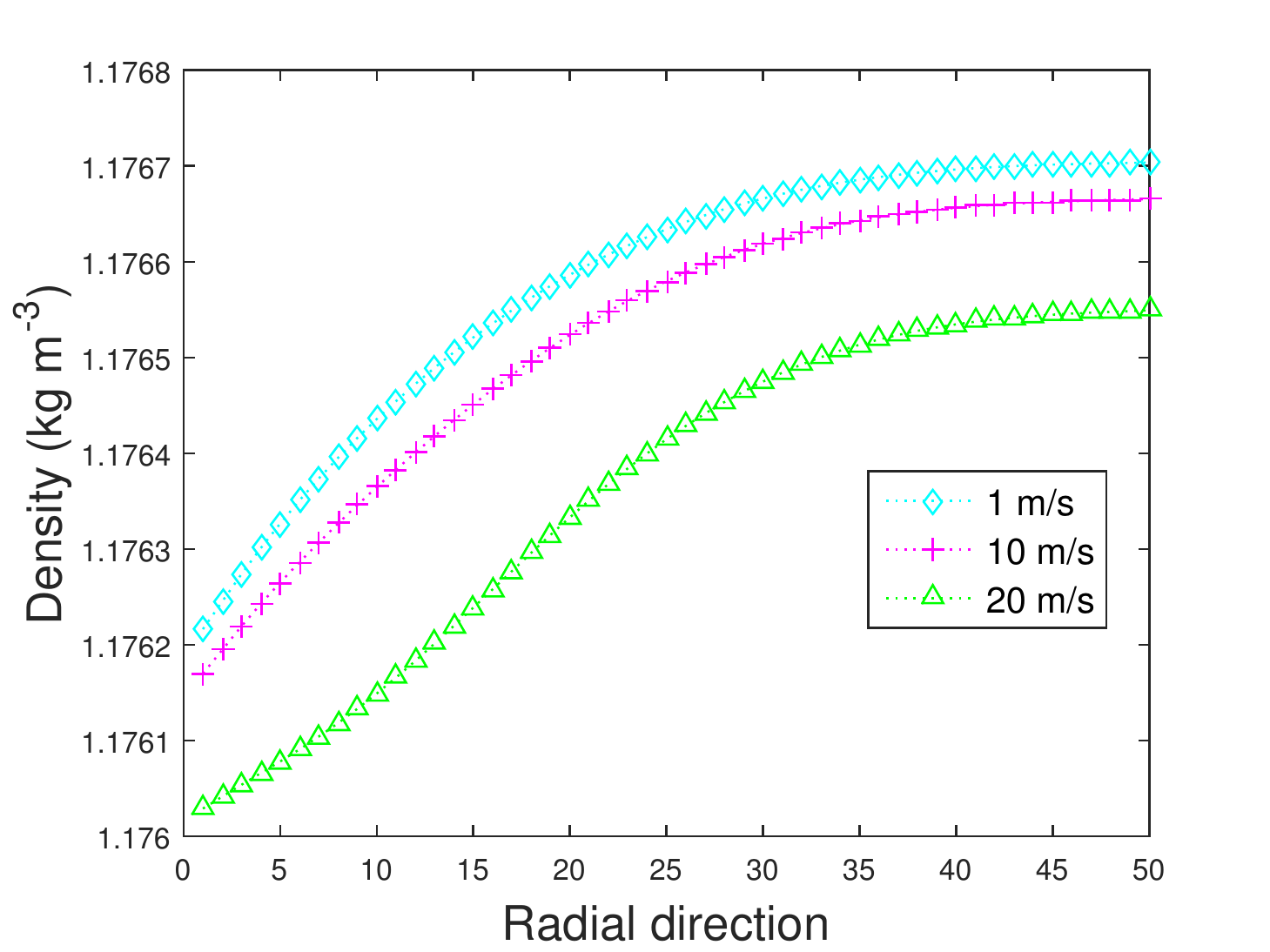}
			
			\caption{Density Distribution at Subsonic Speeds.}
			\label{fig:RhoSub}
		\end{minipage}
		\hspace{1cm}
		\begin{minipage}[b]{0.45\linewidth}
			\centering
			
			\includegraphics[width=9cm]{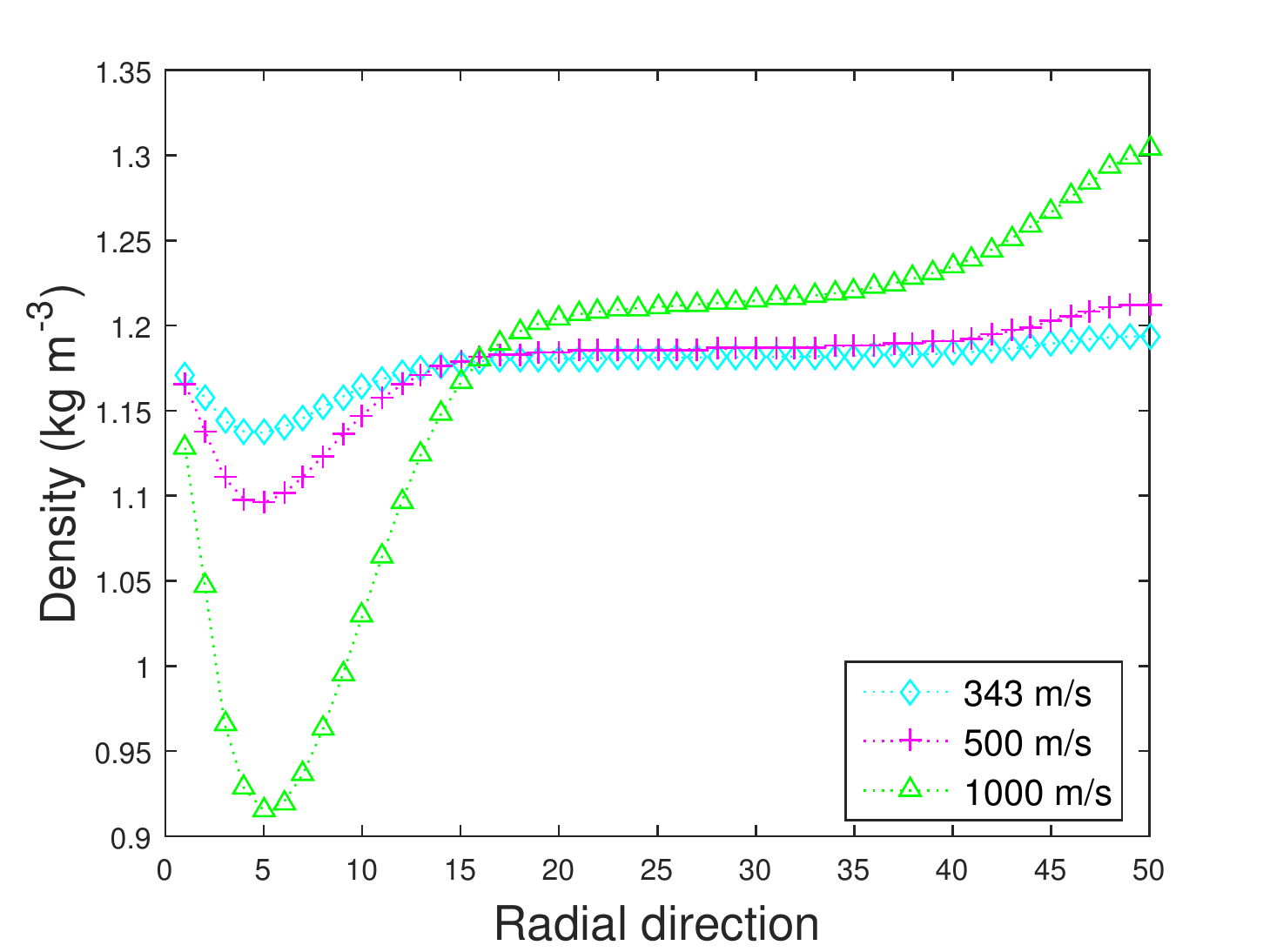}
			
			\caption{Density Distribution at Supersonic Speeds.}
			\label{fig:RhoSuper}
		\end{minipage}
	\end{figure}

    Because high velocities produce low pressures, corresponding density graphs should also be comparatively low. Figures~\ref{fig:RhoSub} and~\ref{fig:RhoSuper} show such consistency. The highest velocity of each graph have the lowest density curve. Most notably, a large dip down to 0.9 $kg/m^3$ is seen in the 1000 m/s case, which reflects the low pressure and high speeds at that point.


	\clearpage
    %
			
			
		%
			
			
	%
	\clearpage

 \section{Future Work}
 The simulation of the flow-field of this cylinder lends to future works well. The flow field of other circular cylindrical objects that spin around its axis can be studied similarly. In particular, the solution to the flow-field can be generated around a Chinese Yoyo, formally known as the Diabolo. The 2-Dimensional Navier-Stokes Equations will be extended to 3 dimensions, and the grid transformation will be from a 3-space to a cylindrical coordinate system. The Diabolo will be modeled by a region of a quadric surface, most likely a hyperboloid of 1 sheet. The finite difference method can still be used, as the general frame of the simulation is the same, but the result should be more interesting.
 
 Another application is to speed up the current simulation. It's implemented in Matlab and has no parallelism. As a result we are restricted to relatively small grid with low resolution. Using Fortran and MPI, the code can be rewritten to run at a much faster rate with a smaller time-step and more time loops. The result will be more accurate, and stability can be achieved for higher resolution grid and higher speed air flow.




\clearpage
\begin{thebibliography}{99}
    \bibitem{Anderson}
	J. D. Anderson, Jr. "Computational Fluid Dynamics, The Basics with Applications," 1995.
	
	\bibitem{Richardson}
	L.F. Richardson, "The Approximate Arithmetical Solution by Finite Differences of Physical Problems." Trans. Roy. Soc. (London), A210:307–357, 1910.
	
	\bibitem{Ghadimi}
	P. Ghadimi and A. Dashtimanesh, "Solution of 2D Navier-Stokes equation by coupled finite difference-dual reciprocity boundary element method," {\it Applied Mathematical Modeling, ScienceDirect}, 2010.
	$<$http://www.elsevier.com/locate/apm$>$
	
	\bibitem{Hirsch}
	C. Hirsch, "Numerical Computation of Internal and External Flows" {\it The Fundamentals of Computational Fluid Dynamics, 2nd Edition}, 1991.
	
	\bibitem{Moukalled}
	F. Moukalled, L. Mangani, and M. Darwish, "The Finite Volume Method in Computational FLuid Dynamics" {\it An Advanced Introduction with OpenFOAM and Matlab}, 2015.
	
	\bibitem{Versteeg}
	H. K. Versteeg and W. Malalasekera, "An Introduction to Computational Fluid Dynamics: The Finite Volume Method," 1995.
	
	\bibitem{Zienkiewicz}
	O. C. Zienkiewicz and R. L. Taylor, "The Finite Element Method, 6th Edition," 2005.
	
	\bibitem{Hosain}
	M. L. Hosain and R. B. Fdhila, "Literature Review of accelerated CFD Simulation Methods towards Online Application," {\it The 7th International Conference on Applied Energy}, 2015.
	
	\bibitem{Matyka}
	M. Matyka, "Solution to two-dimensional Incompressible Navier-Stokes Equations with SIMPLER, SIMPLER and Vorticity-Stream Function Approaches. Driven-Lid Cavity Problem: Solution and Visualization," {\it Computational Physics Section of Theoretical Physics, University of Wroclaw in Poland, Department of Physics and Astronomy}, 2004.
	$<$http://panoramix.ift.uni.wroc.pl/$\sim$maq$>$
	
	\bibitem{Holmes}
	D. G. Holmes and S. D. Connell, "Solution of the 2D Navier-Stokes Equations on Unstructured Adaptive Grids," {\it American Institute of Aeronautics and Astronautics}, 1989.

	\bibitem{Yu}
	D. Yu and A. Kareem, "Two-dimensional Simulation of Flow Around Rectangular Prisms," {\it Journal of Wind Engineering and Industrial Aerodynamics}, vol. 62, pp. 131-161, 1996.
	
	\bibitem{Merrick} 
	R. Merrick and G. Bitsuamlak, "Control of flow around a circular cylinder by the use of surface roughness: A Computational and Experimental Approach," {\it Department of Civil and Environmental Engineering/International Hurricane Research Center, Florida International University}, 2014.
	
	\bibitem{Sobera}
	M. P. Sobera, C. R. Kleijn, H. E. A. van den Akker, and P. Brasser, "Numerical Simulations of the Flow Around a Circular Cylinder Covered by a Porous Medium," {\it Blowing Hot and Cold: Protecting Against Climatic Extremes, RTO HFM Symposium}, 2001. 
	
	\bibitem{Gera}
	B. Gera, P. K. Sharma, and R. K. Singh, "CFD Analysis of 2D unsteady flow around a square cylinder," {\it International Journal of Applied Engineering Research}, vol. 1, no 3, 2010.
	
	\bibitem{Pang}
	A. L. J. Pang, M. Skote, and S. Y. Lim, "Modeling high Re flow around a 2D cylindrical bluff body using the k-$\omega$ (SST) turbulence method," {\it Nanyang Technological University}, vol. 16, no 1, 2016.
	
	
	\bibitem{Maccormack}
	R. W. MacCormack, "The Effect of viscosity in hypervelocity impact cratering," AIAA Paper, pp. 69-354, 1969.

	
	
\end{thebibliography}
\end {document}